%% file: arXiv-version2_26june2026.tex
\documentclass[twocolumn,preprintnumbers,amsmath,amssymb,floatfix,superscriptaddress,longbibliography,prl]{revtex4-1}
\usepackage{siunitx}
\usepackage{graphicx}
\usepackage{float}
\usepackage[caption = false]{subfig}
\usepackage{pgf}
\usepackage{bm}
\usepackage{dsfont}
\usepackage{verbatim}
\usepackage{multirow}
\usepackage{enumitem}
\usepackage{mathrsfs}
\usepackage{leftidx}
\usepackage{xspace}
\usepackage{braket}
\usepackage{bbm}
\usepackage{cancel}
\usepackage{amsfonts,amssymb,amsmath}
\usepackage[normalem]{ulem}
\usepackage{xr}
\usepackage{hyperref}
\hypersetup{colorlinks=true,linktoc=all,linkcolor=blue,breaklinks=true,citecolor=blue,urlcolor=blue}
\usepackage[english]{babel}
\usepackage{cleveref}

\newcommand{\mc}{{\mu_\text{col}}}
\newcommand{\mci}{{\mu_{\text{col},i}}}
\newcommand{\ms}{{\mu_\text{sun}}}
\newcommand{\mm}{\mu^\text{max}_\mathrm{col}}

\newcommand{\Ic}{{I_\text{col}}}
\newcommand{\Ici}{{I_{\text{col},i}}}
\newcommand{\Tc}{{T_\text{col}}}
\newcommand{\Ts}{{T_\text{sun}}}
\newcommand{\Ta}{{T_\text{abs}}}
\newcommand{\Eg}{{E_\text{g}}}
\newcommand{\Em}{{E_0^\text{max}}}

\newcommand{\gs}{{\gamma_\text{sun}}}

\newcommand{\gci}{{\gamma_{\text{col},i}}}

\newcommand{\fa}{{f_\text{abs}}}

\begin{document}
\title{Improving photovoltaics by adding extra terminals to extract hot carriers}

\author{Bruno Bertin-Johannet}
\affiliation{Department of Microtechnology and Nanoscience (MC2),Chalmers University of Technology, S-412 96 G\"oteborg, Sweden}

\author{Thibault Thuégaz}
\affiliation{Department of Microtechnology and Nanoscience (MC2),Chalmers University of Technology, S-412 96 G\"oteborg, Sweden}

\author{Janine Splettstoesser}
\email{janines@chalmers.se}
\affiliation{Department of Microtechnology and Nanoscience (MC2),Chalmers University of Technology, S-412 96 G\"oteborg, Sweden}

\author{Robert S. Whitney}
\email{robert.whitney@grenoble.cnrs.fr}
\affiliation{Laboratoire de Physique et Mod\'elisation des Milieux Condens\'es, Universit\'e Grenoble Alpes and CNRS, B.P. 166, Grenoble 38042, France}

\begin{abstract}
  Photovoltaic cells usually have two terminals, one collecting electrons and the other collecting holes. Can more terminals improve such solar cells? Energy-filtering terminals could collect "hot" carriers (electrons or holes not yet relaxed to the band edge), with other terminals for low-energy carriers. When collection is faster than carrier-phonon relaxation, we predict that four-terminal cells have higher power output than optimal two-terminal cells, often 20-40\% higher. Similar effects will occur in multi-terminal thermoelectrics exploiting non-equilibrium distributions.
\end{abstract}

\date{June 26. 20206}
\maketitle

%%%%%%%%%%%%%%Introduction

\noindent
\textit{Introduction.---} Solar cells produce electrical power because solar photons create electron-hole pairs in an absorber, and the electrons are collected by one terminal (or collector), while the holes are collected by another~\cite{Shockley1961Mar,Wurfel2005}. This then supplies electrical power to a load connected  to the terminals. Numerous ways to increase power outputs have been explored~\cite{Green-book2006,Green2021Jan,Ghasemi2025Jun}. Here we focus on a method known as {\it hot-carrier photovoltaics}, which was proposed 40 years ago \cite{Ross1982May,Wurfel1997Apr,book-chapters}, but only explored in experiments more recently \cite{Clavero2014Feb,Nguyen2018Mar,Chen2020Jun,Konig2020Jun,Fast2021Jun,paul2021hot,fast2022optical,giteau2022hot,viji2024hot,gong2024hot}.  
Unlike conventional photovoltaics, this method aims to get more energy from each charge carrier (electron or hole) by collecting it while it is still approximately as ``hot'' as the thermal photon that excited it (which is at about 6000 Kelvin) \footnote{Typically ``hot'' is used to refer to any high-energy carrier distribution, including nonthermal distributions that do not have a well-defined temperature~\cite{Chen2020Jun,Fast2021Jun,Kumar2024Jul}.}. More precisely, solar photons often excite electrons and holes to energies well above the semiconductor gap, and the aim is to collect the carriers before their interaction with the substrate's phonons (at about 300 Kelvin) causes the carriers to relax to the energy of the gap.  This relaxation causes the carriers to lose a significant part of their useful energy, which becomes heating of phonons and escapes unexploited into the substrate.  To convert all the hot carriers' useful energy into electrical power, one needs an energy-filter on the terminal. This allows the cell to generate a current against a large bias by blocking the backflow of low-energy carriers through the photovoltaic. However, there is a problem with this energy-filtering; it stops the collection of low-energy carriers, such as those excited by low-energy solar photons, or those that have lost part of their energy to the substrate's phonons.
Thus, hot-carrier photovoltaics get more power per carrier collected, but they collect less carriers, making it challenging to achieve an overall higher power output.

%%%%%%%%%%%%%%%%%%%%%%%%%%%%%%%%%%%%%%%%%%%%%%%%%%%%%%%%%%%%%%%%%%%%%%%%%%%%%%%%%%%%%
\begin{figure}[b]
  \begin{center}
    \includegraphics[width=0.98\columnwidth]{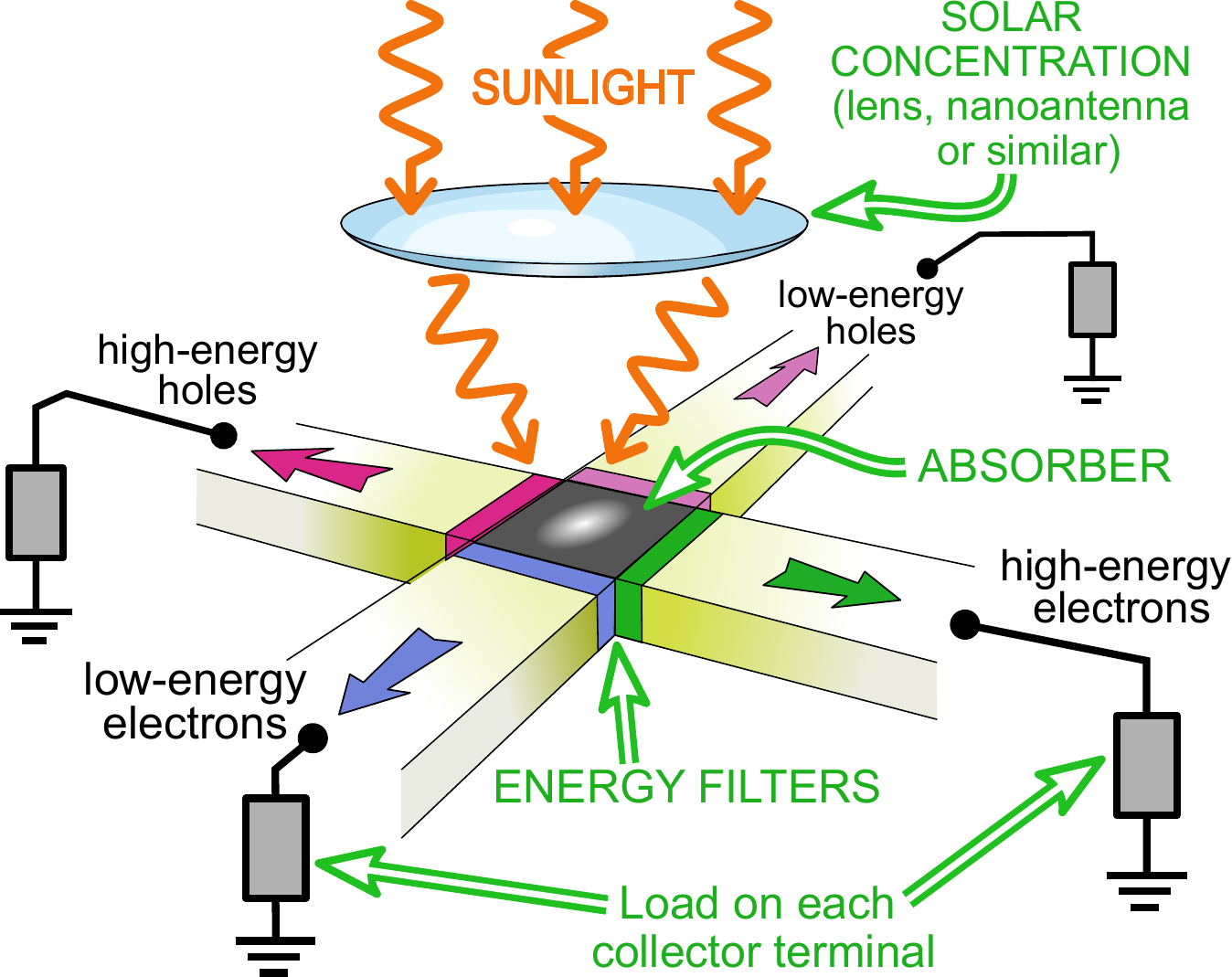}
    \caption{
    The four-terminal version of the multi-terminal photovoltaics proposed here. Sunlight is concentrated on the absorber (by a lens, nanoantenna \cite{Chen2020Jun}, or similar) where it excites electrons and holes. Each terminal's energy-filter  selects the type and energy-interval of carriers that it collects from the absorber, and each terminal is coupled to its own load.}
    \label{fig:setup}
  \end{center}
\end{figure}
%%%%%%%%%%%%%%%%%%%%%%%%%%%%%%%%%%%%%%%%%%%%%%%%%%%%%%%%%%%%%%%%%%%%%%%%%%%%%%%%%%%%%

In this Letter, we propose overcoming this problem by abandoning the conventional approach in which the absorber (where electrons and holes are excited) is coupled to only two terminals---one electron-collecting terminal and one hole-collecting terminal. We propose adding terminals to the absorber, like the four terminals in Fig.~\ref{fig:setup}. Then energy-filtering terminals can collect high-energy carriers to generate a larger amount of useful work per carrier, together with other terminals that collect the lower-energy carriers, allowing them to generate additional power (with a lower useful work per carrier). Each terminal has its own load, tuned for maximal total power output. We first discuss a toy-model in which carrier thermalization is absent,
and show that the maximum power output with a four-terminal setup can be more than 40\% higher than with a two-terminal one.  
We then add the thermalization processes to the model, using an earlier scattering-matrix model \cite{tesser2023} of thermalization between carriers, carrier-phonon thermalization, and electron-hole recombination. 
These processes combine with the carrier extraction to create a nonthermal distribution of carriers in the absorber.
It shows that two-terminal power outputs are largest in the ``hot-carrier'' limit \cite{Ross1982May,Wurfel1997Apr}, where thermalization between carriers is
much faster than the other processes listed above; then we find that adding terminals gives only small improvements.
However, this limit is hard to access experimentally. Experiments often show other processes occurring at similar rates to thermalization between carriers. Then our model predicts that adding terminals can give a more than 20\% improvement in power output.

Our model is simple enough for easy numerical maximization of the power output over a very wide range of parameters. For this, it simplifies carrier generation and recombination, and assumes carrier-carrier and carrier-phonon interactions are energy-independent. None the less, its predictions indicate promising regimes for adding terminals. These regimes should then be explored experimentally, or using heavy numerics for precise microscopic modeling (Green's function \cite{Michelini2017Mar,aram2017quantum, cavassilas2022theoretical,dai2023recent}, density-matrix~\cite{Mueller2018Sep} or ab initio \cite{roknuzzaman2017towards,solet2025many} techniques, etc). 
Similarly, data from such experiments or numerics would provide parameters to make our model more realistic.

Our model also predicts that adding many more terminals (each with a chosen energy-filter and load) leads to further increases in power output, but with the biggest increase occurs when going from two to four terminals, see End Matter. However, the limit of an infinite number of terminals, coupled to an absorber small enough that carriers do not thermalize, shows a power output that is 80\% of the incoming solar heat flow.

%%%%%%%%%%% Method

\vskip 2mm \noindent
\textit{Method.---} We use a scattering matrix approach~\cite{Moskalets2011Sep,Blanter2000Sep,tesser2023} to model our proposed multi-terminal photovoltaic.
Multi-terminal devices have been widely considered for versatile and tunable thermoelectric effects, see e.g.~Refs.~\cite{Entin-Wohlman2010Sep,Sanchez2011Feb,Sothmann2014Aug,Mazza2014Aug,Whitney2016May,Benenti2017Jun}, and exploiting non-thermal resources~\cite{Sanchez2019Nov,Hajiloo2020Oct,Sanchez2019Oct,Ciliberto2020Nov,Freitas2021Mar,Deghi2020Jul}. However, we do not know of any studies of power outputs as one increases the number of terminals, 
neither for thermoelectrics nor for photovoltaics.

Here, for simplicity, we assume the electron and hole dynamics are the same (same masses, same interactions, etc),
and that for every terminal collecting electrons, there is an identical one for holes.
Then we need only maximize the electron contribution to power output, and double it to get the full power output. %(extension to asymmetric cases are straightforward).
Using a convention with $\hbar=k_\text{B}=|e|=1$, the particle current from the absorber to collector terminal $i$ is 
\begin{equation}\label{eq:current-scattering}
  \begin{aligned}
    \Ici & =
    \frac{1}{2\pi}\int_{0}^{\infty}\mathrm{d}\varepsilon\, D_i(\varepsilon)
    \left\lbrack \fa(\varepsilon)- f_{\text{col},i}(\varepsilon)
    \right\rbrack\, ,
  \end{aligned}
\end{equation}
where $D_i(\varepsilon)$ is the transmission probability for carriers at energy $\varepsilon$ to pass between the absorber and terminal $i$.
The integral starts in the middle of the absorber's band gap, $\mu_\mathrm{ref}\equiv0$, so negative energies contribute to hole currents. Terminal $i$ has a Fermi distribution $f_{{\rm col},i}(\varepsilon)=(1+\exp[(\varepsilon-\mu_i)/\Tc])^{-1}$ for room temperature ($\Tc=300\,$K), and that terminal's load $R^{\rm load}_i$ is determines its electrochemical potential, $\mci = \Ici R^{\rm load}_i$.
The absorber's distribution, $\fa(\varepsilon)$, is self-consistently determined by all the processes contributing to carrier dynamics, it is usually a nonequilibrium distribution, and may be a non-monotonic function of $\varepsilon$.
The power output into terminal $i$ is a current $I_i$ {\it against} a potential $\mu_i$
\begin{eqnarray}
    P_{\mathrm{col},i} =  \Ici \,\mci  = I_{{\rm col},i}^2 R^{\rm load}_i\ .
    \label{Eq:power-output-def}
\end{eqnarray}
We tune each $R^{\rm load}_i$ to maximize the total power output $\sum_i P_{\mathrm{col},i}$.
This power output is typically compared to the power of the incoming heat flow from the sun. For this, we assume a perfect absorber, where every solar photon creates an electron and a hole in a thermal distribution at $T_\mathrm{sun}=6000\,$K, that is available to extract (see End Matter), so the incoming heat flow is 
\begin{eqnarray}
    P_\mathrm{sun}^\mathrm{in} =  \frac{\gamma_\mathrm{sun}}{2\pi} \int_0^\infty \frac{\varepsilon\, d\varepsilon }{1+\exp[\varepsilon/\Ts]}    = 4.7\times 10^6 \,\gamma_\mathrm{sun} \, .
    \label{Eq:power-sun-in}
\end{eqnarray}
where $\gamma_\mathrm{sun}$ is a dimensionless number adjusted to fit the rate at which solar energy arrives at the absorber \footnote{Multiplying by $k^2_{\rm B}/\hbar$ to get SI units, gives $ P_\mathrm{sun}^\mathrm{in} = 8.5\, \mu$W at $\gamma_\mathrm{sun}=1$, equivalent to sunlight focused onto the absorber by a lens of about 0.1\,mm diameter}.

The effect of adding terminals is most easily understood for what we call a {\it thermoelectric toy-model}, so we first explain that. Our End Matter also explains it for the well-known {\it ultimate efficiency}  model of Shockley and Queisser \cite{Shockley1961Mar}.
We then make our model realistic by adding both carrier-phonon losses and thermalization between carriers. We explore the hot-carrier effect of Refs.~\cite{Ross1982May,Wurfel1997Apr} within this model. There the thermalization between carriers can boost the two-terminal power output, but we show that adding terminals can give a significant further improvement in power output.

%%%%%%%%%%%%%%%%%%%%%%%%%%%%%%%%%%%%%%%%%%%%%%%%%%%%%%%%%%%%%%%%%%%%%%%%
\begin{figure}[tb]
  \subfloat{\centering\includegraphics[width=.9\linewidth,
  keepaspectratio]{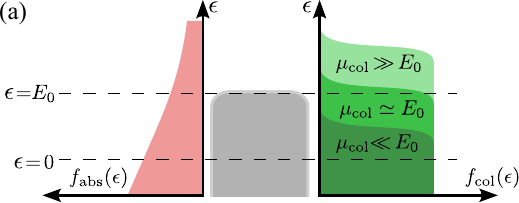}}\\
\subfloat{\centering\includegraphics[width=.85\linewidth,
  keepaspectratio]{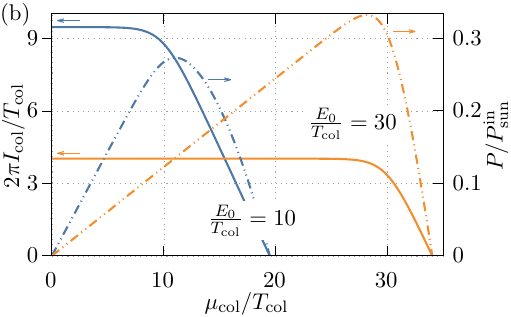}}
  \caption{(a) Energy landscape of the two-terminal setup for three
    different collector chemical potentials $\mu_\text{col}$---corresponding to three different loads. (b) The current (full lines) and the
    power (dashed lines) for two different filtering energies, \(E_0/\Tc =
    10, 30\), as a function of \(\mc\).
    %, obtained from $\mc = R^\text{load} I_\text{col}$. 
    }
  \label{fig:thermoelectric_2terminal}
\end{figure}
%%%%%%%%%%%%%%%%%%%%%%%%%%%%%%%%%%%%%%%%%%%%%%%%

%%%%%%%%%%%%%%%%%%%%%%%%%%%%%%%%%%%%%
\vskip 2mm \noindent
\textit{Toy-model with two-terminals.---}
We start by recalling how an energy filter can be a two-terminal photovoltaic. We do this with a toy-model where carriers move without thermalizing, but that any carrier that does not have the right energy to get through the energy-filter eventually undergoes recombination that re-radiates its energy back to the sun \cite{tesser2023,Tesser2022}.
Hence, this toy-model assumes that solar photons induce a thermal distribution of carriers in the absorber with the solar spectrum's temperature ($\Ta =\Ts = 6000\,$K, $\mu_{\rm abs}=0$), so $f_{\rm abs}(\varepsilon)=1/(1+\exp[\varepsilon/\Ts])$.
Then an energy-filter acts like a nanoscale thermoelectric (see Ref.~\cite{Benenti2017Jun} for a review) between two terminals at
$\Ts=6000\,$K and $\Tc=300\,$K, so we call it a {\it thermoelectric toy-model} of a photovoltaic.

We take a step-function energy-filter between absorber and collector terminal (shown to be optimal for power production in \cite{whitney2015,Whitney2016May,Tesser2022}) such that only electrons above energy \(E_0\) can pass. Then $D(\varepsilon< E_0) = 0$ and  $D(\varepsilon> E_0) = 1$,
and the current
$2\pi \Ic/\gamma_{\rm sun} = -\Ta\ln\left[f_\text{abs}(E_0)\right]
 + \Tc \ln\left[f_\text{col}(E_0)\right]$.
% \begin{equation}\label{eq:cur-two-cont-fil}
%   2\pi \Ic = -\Ta\ln\left[f_\text{abs}(E_0)\right]
%   + \Tc \ln\left[f_\text{col}(E_0)\right] \, .
% \end{equation}
% The current and power output are shown in Fig.~\ref{fig:thermoelectric_2terminal}(b) for \(E_0=10\Tc\) and \(E_0=30\Tc\).
 Fig.~\ref{fig:thermoelectric_2terminal}(a) shows this for three regimes of $(\mc-E_0)/\Tc$ from dark to light green. The first regime has a low load resistance, so $(\mc-E_0)/\Tc\ll0$. Then hot carriers flow from absorber to collector, with negligible back-flow since collector electrons have too little energy to get over the barrier. 
Then the 
%current is $\mc$-independent and larger for lower $E_0$, so 
power output grows by reducing $E_0$ or increasing the load resistance to raise $\mc$. This pushes us into a second regime where $(\mc-E_0)/\Tc\sim0$. It has a backflow of electrons from collector to absorber, adding a negative contribution to the power output that becomes stronger as $\mc$ increases or $E_0$ decreases. Finally, the third regime is when the load resistance is large enough for $(\mc-E_0)/\Tc \gg0$. Then the collector's distribution exceeds the absorber's in a growing energy interval above $E_0$, so there is a strong backflow, causing the linear decay of $\Ic$ with $\mc$ in Fig.~\ref{fig:thermoelectric_2terminal}(b). $\Ic$ vanishes when $\mc$ is the  \textit{stopping
voltage} (infinite load resistance).
%at which the thermo-current is canceled by the backflow. 
Raising $E_0$ raises the optimal $\mc$, but lowers the optimal $\Ic$, and the power output (dashed lines Fig.~\ref{fig:thermoelectric_2terminal}) is $\mc \Ic$. Hence, there is a clear
trade-off between extracting many low-energy electrons or fewer high-energy electrons.

Ref.~\cite{whitney2015}'s analytic method to find the \(E_0\) and
\(\mc\) that maximize power output, yields $f_\text{abs}(\Em)=f_\text{col}(\Em)$, so the filter blocks energies with a net flow in the wrong direction.
%, that would otherwise occur at all $\varepsilon$ with $f_\text{abs}(\varepsilon)<f_\text{col}(\varepsilon)$. 
Then using $x\ln(x)+(x+1)\ln(x+1)=0$ \cite{whitney2015},
where $x=\exp\left[-\mm/(\Ta-\Tc)\right]$, yields
maximum power at $\mm=21.7\Tc$ and $\Em=22.89\Tc$.
Hence,  the maximum two-terminal power output is 
\begin{eqnarray}
P_\textrm{toy-model}^\textrm{2-term\,max}= 0.34 \,P_\mathrm{sun}^\mathrm{in},
\label{Eq:P_2term-ideal}
\end{eqnarray} 
for the thermoelectric toy-model considered here.

%%%%%%%%%%%%%%%%%%%%%%%%%%%%%%%%%%%%%

\vskip 2mm \noindent
\textit{Toy-model with four terminals.---} We now add terminals to the thermoelectric toy-model to circumvent the above tradeoff between extracting many electrons at
low energy or fewer electrons at high energy.
Energy filters ensure that high-energy electrons can enter terminal A  while low-energy electrons can enter terminal B. 
We take boxcar-shaped filters \footnote{Boxcar-shapes could be implemented with narrow bandgap materials \cite{zhou2011optimal} or nanostructures \cite{whitney2015}},
\begin{subequations}\label{eq:2transmissions}
\begin{eqnarray}
  D_\text{A}(\varepsilon) & = & 
  \begin{cases}
    1 & \text{ if}\ E_{\rm A0} < \varepsilon < E_{\rm A1}\\
    0 & \text{ otherwise}\, ,
  \end{cases}\\
  D_\text{B}(\varepsilon) & = & 
  \begin{cases}
    1 & \text{ if}\ E_{\rm B0} < \varepsilon < E_
    {\rm B1}\\
    0 & \text{ otherwise}\, .
  \end{cases}
\end{eqnarray}
\end{subequations}
and we assume $\Eg < E_{\rm B0} < E_{\rm B1} <E_{\rm A0} < E_{\rm A1}$ where the absorber's semiconducting gap is $\Eg$.

For the above thermoelectric toy-model (we address more realistic models below), the physics of adding terminals is fairly straightforward, however
a small numerical optimization is required to give the maximum power output $P_\mathrm{tot}$.
Similarly, the physics of adding terminals is fairly straightforward for a well-known model of Shockley and Queisser \cite{Shockley1961Mar}, as explained in our End Matter.

\begin{figure}[tb]
  \includegraphics[width=.8\linewidth,
  keepaspectratio]{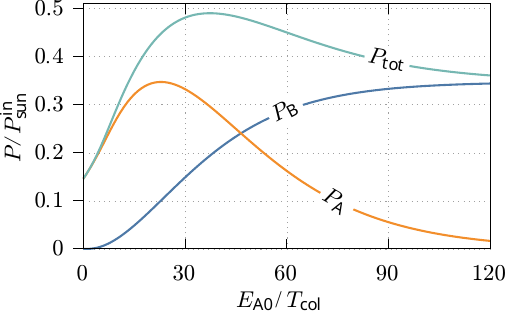}
  \caption{
  Maximum total power output $P_\mathrm{tot}$, for the thermoelectric toy-model with four-terminals. We plot $P_\mathrm{tot}$ versus $E_{\rm A0}$, when $E_{\rm A1}\to \infty$ (its optimal value in this toy-model), and we have numerically optimized $E_{\rm B0}$, $E_{\rm B1}$, and collector chemical potential $\mu_\mathrm{A},\mu_\mathrm{B}$  (equivalent to optimizing the loads $R^\text{load}_\text{A}$, $R^\text{load}_\text{B}$). 
  For each $E_{\rm A0}$, we show the separate components $P_\mathrm{A},P_\mathrm{B}$ of this maximized $P_\mathrm{tot}$.}
  \label{fig:thermoelectric_multi}
\end{figure}

Fig.~\ref{fig:thermoelectric_multi} shows the optimization of our thermoelectric toy-model for varying $E_{\rm A0}$, when optimized over $E_{\rm B0}$, $E_{\rm B1}$, $E_{\rm A1}$ and collector loads $R^\text{load}_\text{A}$, $R^\text{load}_\text{B}$ (here optimal $E_{\rm A1}\to\infty$). 
The higher energy collector A is similar to the two-terminal case; its power \(P_\text{A}\) (orange line in Fig.~\ref{fig:thermoelectric_multi}) is maximum at \(E_{\rm A0} =22.89 \Tc\). 
The other collector B has power $P_\mathrm{B}$ (blue line) that increases with
\(E_{\rm A0}\), and dominates once \(E_{\rm A0}\) is too high for carriers to enter A.
At optimal $E_{\rm A0}$, 
\begin{eqnarray}
P_\textrm{toy-model}^\textrm{4-term\,max}= 1.41 \,
P_\textrm{toy-model}^\textrm{2-term\,max},
\label{Eq:P_4term-ideal}
\end{eqnarray}
So this thermoelectric toy-model gets about 40\% more power output with four-terminal than with two-terminals in Eq.~(\ref{Eq:P_2term-ideal}). This is a conversion of 48\% of the sun's heat energy into electrical power.
Smoother shaped $D_{\rm A,B}(\varepsilon)$ than Eq.~(\ref{eq:2transmissions})'s boxcars are considered in our End Matter. While less good than boxcars,  they still give a significant improvement over two-terminal devices.

%%%%%%%%%%%%%%%%%%%%%%%%%%%%%%%%%%%%%%%%%%%%%%%%%%%%%%%%%%%%%%%%%%%%%%%%
%%%%%%%%%%%%%%%%%%%%%%%%%%%%%%%%%%%%%%%%%%%%%%%%%%%%%%%

\vskip 2mm \noindent
\textit{Toy-model with infinitely many terminals.---}  Our End Matter's Fig.~\ref{fig:inf-filt} shows that adding even more terminals to the thermoelectric toy-model  further increases in power output, but that the improvement becomes less the more terminals one adds. A crude fit gives $P(N) \simeq P_\textrm{toy-model}^{N\to\infty \,\rm{max}}\times(1 - \frac{1}{N+1}) $ for $N$ electron terminals (so a four-terminal device has $N=2$), with the $N\to \infty$ limit being $P_\textrm{toy-model}^{N\to\infty \,\rm{max}}\simeq 0.8 P_\mathrm{sun}^\mathrm{in}=2.35 P_\textrm{toy-model}^\textrm{2-term max}$. Thus, an infinite number of terminals converts 80\% of the sun's heat energy into electrical power, more than double that of the same toy-model with two-terminals.

%%%%%%%%%%%%%%%%%%%%%%%%%%%%%%%%%%%%%
\begin{figure}
\includegraphics[width=0.95\columnwidth]{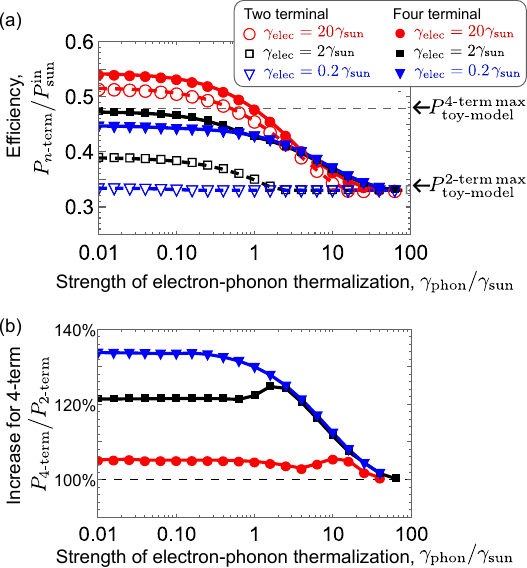}
  \caption{(a) Maximized power in the presence of carrier thermalization effects, plotted versus the strength of losses due to electron-phonon thermalization. Different colors are for carrier-carrier thermalization that is weak (blue triangles), medium (black squares) or strong (red dots), $\gamma_{\rm elec}=\{0.2,2,20\}$. In all cases, recombination is weak, $\gamma_{\rm rec}=0.01$.
  For four-terminal devices (solid symbols), we assume $\gamma_{\rm col\,A}+\gamma_{\rm col\,B}=20\gamma_{\rm sun}$, and maximize power over the eight remaining parameters; these are $\gamma_{\rm col\,A}$, $\Eg$ (semiconducting gap), $E_{\rm B0}< E_{\rm B1} <E_{\rm A0} <E_{\rm A1}$ (energies of boxcar filters), $\mu_{\rm A}$ and $\mu_{\rm B}$ (equivalent to optimizing each terminal's load). 
  For two-terminal devices (open symbols), we have no terminal B, we assume $\gamma_{\rm col}=20\gamma_{\rm sun}$, and maximize power over the four remaining parameters ($\mu_{\rm A}$ and $\Eg <E_{\rm A0}<E_{\rm A1}$). 
    (b) Results in (a) presented as the ratio of maximized power for four-terminals 
    %(two collector pairs) 
    to maximized power for two-terminals. 
    %(one collector pair)
    }
  \label{fig:neq}
\end{figure}
%%%%%%%%%%%%%%%%%%%%%%%%%%%%%%%%%%%%%
%

%%%%%%%%%%%%%%%%%%%%%%%%%%%%%%%%%%%%%
\vskip 2mm \noindent
\textit{Full model including thermalization.---} We now add thermalization processes, specifically carrier-carrier interactions, carrier-phonon interactions, and electron-hole recombinations. This complicates the analysis, but it captures the hot-carrier effect, along with realistic carrier distributions. 
For this, we use  a multi-terminal version of Ref.~\cite{tesser2023}'s model, in which
interaction and recombination processes are each modeled as a B\"uttiker probe~\cite{Buttiker1985Aug,Buttiker1986Mar,Sanchez2011Nov}. This is a nanoscale equivalent of the relaxation-time approximation in Boltzmann transport equations (see e.g., Ref.~\cite{Brennan1999-semiconductor-book}), so it neglects the energy-dependence of each process, but its strength could be fitted from experiments or microscopic modeling.  

The absorber's carrier distribution is a weighted average of the probe distributions (see End Matter),
  $\fa(\varepsilon) =
  \sum_i
  \gamma_i (\varepsilon)
  f_i(\varepsilon)\big/\sum_i
  \gamma_i (\varepsilon),$
% \begin{equation}\label{eq:fabs}
%   \fa(\varepsilon) =
%   \frac{\sum_i
%   \gamma_i (\varepsilon)
%   f_i(\varepsilon)}{\sum_i
%   \gamma_i (\varepsilon)}\,
%   ,
% \end{equation}
with $i$ being summed over six distributions
$i\in\{$sun, phon,  rec, col$_\mathrm{A}$, col$_\mathrm{B}$, elec$\}$.
The $\gamma_i$s are dimensionless parameters giving relative rates, with $\gamma_\mathrm{sun}$ explained below Eq.~(\ref{Eq:power-sun-in}).
The distribution $\fa(\varepsilon)$ is often non-trivial and non-monotonic, since
the flow into a collector can cause it to dip at energies where this flow occurs  (see End Matter).  This complexity implies that no simple arguments can give ideal filter parameters, and a numerical optimization is required.

Figure~\ref{fig:neq} shows results of a numerical maximization
of total power output for two and four terminals 
for certain values of $\gamma_{\rm elec}/\gamma_{\rm sun}$ and  $\gamma_{\rm ph}/\gamma_{\rm sun}$, with weak carrier recombination $\gamma_{\rm rec}/\gamma_{\rm sun}=0.01$.
In each case, the total coupling to collectors is $20\gamma_{\rm sun}$, so a two-terminal case has $\gamma_{\rm col}=20\gamma_{\rm sun}$, while a four-terminal case has $\gamma_{\rm colA}+\gamma_{\rm colB}=20\gamma_{\rm sun}$. We  maximize this power output over all other parameters (listed in Fig.~\ref{fig:neq}'s caption). 
Fast thermalization between carriers, $\gamma_{\rm elec} \geq 20\gamma_{\rm sun}$, gives the  hot-carrier limit of Ref.~\cite{Ross1982May,Wurfel1997Apr}; then the power output of the two-terminal device is already so large that improvements coming from adding terminals are small.
For more modest
thermalization between carriers, $\gamma_{\rm elec}=2\gamma_{\rm sun}$, the thermalization still makes the two-terminal device significantly better than in the absence of thermalization (represented by our thermoelectric toy-model), yet
four-terminal devices outperform the optimum two-terminal devices by 20\% or more.
This improvement only ceases if electron-phonon thermalization causes carriers to relax to the band edge before they are collected.

%%%%%%%%%%%%%%%%%%%%%%%%%%%%%%%%%%%%%%%%
\vskip 2mm \noindent
{\it Conclusions.---}
Despite forty years of effort, 
Refs.~\cite{Ross1982May,Wurfel1997Apr}'s 
{\it hot-carrier} route to high photovoltaic power output has been impeded
by the experimental difficulty of getting
thermalization between carriers to be much faster than other processes
(even if recent progress has been made \cite{Fast2021Jun}). 
We propose a new route to improving power output of hot-carrier solar cells by adding extra terminals, so low- and high-energy carriers are filtered into different collector terminals with different loads. 
Our model shows that this can increase power outputs by 20-40\%. Of particular note are cases where two-terminal power outputs are already boosted by modest thermalization between carriers, then adding terminals could give an extra boost of more than 20\%.

Our model aims to realistically capture the carrier extraction, includes carrier backflow from terminals, and the impact of this extraction on the carrier distribution in the absorber. In contrast, it treats carrier creation and thermalization more simply than many other works (but it obeys thermodynamic laws \cite{Whitney2013Mar}). 
As such, the above percentages are not directly comparable to other models, such as the Shockley-Queisser model (see End Matter). However, we claim that the trend is clear; any model that includes realistic carrier extraction will show improved power output if extra terminals are added.

We expect this multi-terminal strategy may work beyond photovoltaics, and apply to other nanoscale devices  that produce useful work from heat or waste energy in non-equilibrium distributions (including thermoelectrics, photoelectrics \cite{jacob2025thermodynamics}, near-field thermophotovoltaics \cite{mittapally2021near}, etc). 
As many experiments on hot-carrier cells use energy-filters in nanowires~\cite{Limpert2017Oct,Fast2020Jul,fast2022optical}, we mention realizations of multi-terminal devices with nanowires~\cite{Yoon2010Aug,Dick2004Jun,Kumar2024Jul} and their assembly in large arrays~\cite{Wallentin2013Jan,daCamaraSantaClaraGomes2024May} as promising for implementing our proposed multi-terminal devices.

%%%%%%%%%%%%%%%%%%%%%%%%%%%%%%%%%%%%%%%%
\vskip 2mm \noindent
\textit{Code and Data.---} Code and data for all figures are at 
\href{https://doi.org/10.5281/zenodo.16994493}{DOI:10.5281/zenodo.16994493}, except the final version of Fig.~\ref{fig:neq} whose code and data are at 
\href{https://doi.org/10.5281/zenodo.18702207}{DOI:10.5281/zenodo.18702207}.
%%%%%%%%%%%%%%%%%%%%%%%%%%%%%%%%%%%%%%%%
\vskip 2mm \noindent
\textit{Acknowledgments.---}
%%%%%%%%%%%%%%%%%%%%%%%%%%%%%%%%%%%%%%%%
We thank Ludovico Tesser, Elsa Danielsson, and \mbox{Krishna} Lyn Delima for helpful discussions. We were encouraged in this work by Adam Burke and Heiner Linke. Funding from the Knut and Alice Wallenberg Foundation via the Fellowship program (B.B-J.,J.S.) and  from the European Research Council (ERC) under the European Union’s Horizon Europe research and innovation program (101088169/NanoRecycle) (J.S.) is gratefully acknowledged.
Furthermore, R.W. acknowledges the support of the French National Research Agency (ANR)
through the project ``TQT''(ANR-20-CE30-0028),
the project ``QuRes'' (ANR-21-CE47-0019), and 
the OECQ project that is financed by the French state (via France 2030) and Next Generation EU (via France Relance).

\newpage

%%%%%%%%%%%%%%%%%%%%%%%%%%%%%%%%%%%%%%%%
\section*{End Matter}
\appendix
%%%%%%%%%%%%%%%%%%%%%%%%%%%%%%%%%%%%%%%%

\noindent
\textit{Adding terminals to the Shockley-Queisser model}.---
To show how adding terminals can improve efficiency in other models of photovoltaics, we take Shockley and Queisser's simple {\it ultimate efficiency model} in section 2 of Ref.~\cite{Shockley1961Mar}. 
For a gap $E_g$, they assume that any photon absorbed at energy $E \geq E_g$ creates an electron and hole that relax to their band edges, after which their total energy is $E_g$ (with all other energy lost to phonons). By taking this as the {\it only} thermalization mechanism, and so assuming all the remaining carrier energy is perfectly converted into useful work (as electrical power), they get an upper-limit on efficiency for a two-terminal photovoltaic,
$u(x_g) =  x_g \int_{x_g}^\infty  n(x)\,dx \Big/\int_0^\infty x\, n(x)\,dx,$
% \begin{eqnarray}
% u(x_g) =  \frac{x_g \int_{x_g}^\infty  n(x)\,dx}{\int_0^\infty x\, n(x)\,dx}\ ,
% \end{eqnarray}
where $n(x)=x^2\big/(e^x-1)$ and $x=E/(k_{\rm B} \Ts)$. Here, the denominator gives the energy arriving from solar photons (blackbody radiation at $\Ts$) and the numerator gives the energy in the carriers after their relaxation to the band edge. 
The maximum of $u(x_g)$ \cite{Shockley1961Mar} is their {\it ultimate limit on efficiency} (for two-terminals) of about 44\%.

Now we add a pair of terminals with energy-filters at $E_1$, and assume they extract 
all carriers created by photons with energy $E$, such that $E>E_1>E_g$, before those carriers relax.  
Meanwhile, photons with energies between $E_g$ and $E_1$ create carriers that relax to the band edge and are extracted to the other terminals.
In the same spirit as the above {\it ultimate efficiency model},  we assume the carriers entering the new energy-filtering terminals relax to energy $E_1$, but that energy is then available to be turned into useful work. 
%(Our microscopic model of extraction show that a small part of this energy is actually lost when converting it into useful work, since the contact's chemical potential must be below $E_1$ to avoid a back-flow of carriers from the contact, but we neglect that small loss effect here).  
Then the limit on efficiency with this extra pair of terminals is 
\begin{eqnarray}
u_\text{4-term}(x_g,x_1) =  \frac{x_g \int_{x_g}^{x_1} n(x)\, dx
+ x_1\int_{x_1}^\infty n(x)\, dx}{\int_0^\infty x\, n(x)\,dx},\ \  
\end{eqnarray} 
whose maximum is about 60\%. 
So going from two terminals to four terminals raises the limit on efficiency from 44\% to 60\% --- an increase of 36\%. 

Of course, this model neglects various thermalization processes and some aspects of carrier-extraction (like back-flow from the terminals), which are included in our full model. However, this simple calculation reinforces our claim that adding terminals to any given model (with given thermalization processes) will improve its efficiency, and that improvement can be  significant. This strongly suggests that adding terminals to experimental photovoltaics could give similar efficiency improvements.

%%%%%%%%%%%%%%%%%%%%%%%%%%%%%%%%%%%%%%%%

%%%%%%%%%%%%%%%%%%%%%%%%%%%%%%%%%%%%%%%%
\begin{figure}
  \includegraphics[width=0.85\linewidth,
    height=\textheight,
  keepaspectratio]{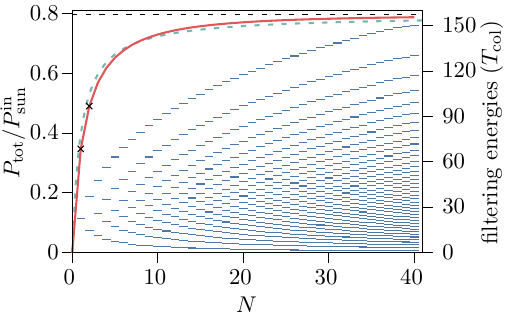}
  \caption{The red curve is the optimal value of the total power output for our thermoelectric toy-model with $N$ collector terminals, when  Eq.~\eqref{eq:Nfilters} gives each terminal's energy filters. It is fairly well fitted by $P(N) =  P_\textrm{toy-model}^{N\to\infty \,\rm{max}}\times(1 - \frac{1}{N+1})$ (dashed green curve). For the red curve, each terminal's load and filter position was optimized at each $N$. Fine blue dashes are the optimal position of each filter's lower edge $(E_\alpha-\delta \epsilon)$. The crosses are the maximal power for the toy-model with two- and four-terminals. The horizontal dashed black line is the power for an infinite number of filters in Eq.~(\ref{eq:Pmax}). }
  \label{fig:inf-filt}
\end{figure}
%%%%%%%%%%%%%%%%%%%%%%%%%%%%%%%%%%%%%%%%

%%%%%%%%%%%%%%%%%%%%%%%%%%%%%%%%%%%%%%%%
\vskip 2mm
\noindent
\textit{Toy-model: Adding more terminals.---}
The main text showed that adding a second pair of terminals with an energy filter can significantly increase the  power generated  by a solar cell described by our thermoelectric toy-model. The question arises; is adding even more terminals beneficial? 
Hence, we analyze a situation where $N$ collector pairs are connected to the absorber with thin, energetically adjacent boxcar-shaped energy filters, taking the thermoelectric toy-model in which $\fa=f_\mathrm{sun}$. 
Each collector $\alpha=1,...,N-1$ is connected to the absorber by an energy filter with width $2\delta\varepsilon$ centered around energy
$E_{\alpha}$. The transmissions hence read
\begin{equation}\label{eq:Nfilters}
  D_{\text{sun},\alpha}(\varepsilon) = 1\ \ \text{  for  }E_\alpha -
  \delta\varepsilon < \varepsilon < E_\alpha + \delta\varepsilon
\end{equation}
and zero otherwise. The $N$th filter is a step function that transmits all carriers with energies above $E_{N-1}+\delta\varepsilon$. 
We now optimize the power output by adjusting loads and filter positions. Fig.~\ref{fig:inf-filt} shows the maximum power as well as the optimal filter positions as a function of the number of
filters, $N$.
Three main conclusions can be drawn from this figure. First,
it is possible
to significantly enhance the power output by adding filters. Second,
only the first few
filters add a substantial contribution to the optimal power output. Third, in order to achieve the large enhancement in the multi-filter case, the placement
of these filters in energy must be optimized.  We
see that it is advantageous to stack filters more densely at lower energies, where the state have higher occupations in the thermoelectric toy-model.
This would pose a substantial experimental challenge in the limit of many terminals.

Now we take the limit of infinitely many terminals, with non-overlapping energy-filters, so each filter's width $\delta\varepsilon \to 0$. Then the current into collector terminal $\alpha$ is
$I_{\mathrm{col},\alpha}  \approx  \left[
      \fa(E_{\alpha}) -
    f_{\mathrm{col},\alpha}(E_\alpha) \right] \delta\varepsilon +
    {\cal O}\left( \delta\varepsilon^{2} \right)$.
The power $P_\alpha=I_{\mathrm{col},\alpha}\mu_\alpha$ is maximized when \(\mu_{\alpha}\) satisfies
$\mu_{\alpha}f'_{\alpha}\left( E_{\alpha} -
\mu_{\alpha} \right) + f_{\alpha}\left( E_{\alpha} -
\mu_{\alpha} \right) = f_{\text{abs }}\left( E_{\alpha} \right)$,
where the prime is the energy derivative.
Then, the total power output (summed over all collectors) is
$P_\textrm{toy-model}^{N\to\infty \,\rm{max}}  = 
\sum_\alpha(\mu_{\alpha}^\text{max})^{2}f'_{\alpha}\left( E_{\alpha} -
  \mu_{\alpha}^\text{max} \right) \delta\varepsilon$. 
Writing the sum as an integral (due to $\delta \varepsilon\to 0$), gives 
\begin{eqnarray}
P_\textrm{toy-model}^{N\to\infty \,\rm{max}} & = & \int d\varepsilon\, \left(\mu^{\text{max}}(\varepsilon)\right)^{2}f'(\varepsilon -
  \mu^{\text{max}}(\varepsilon)), \ \ 
  \label{eq:Pmax}
\end{eqnarray}
This is the horizontal dashed black line in Fig.~\ref{fig:inf-filt}, which the red curve quickly approaches as $N$ increases.

%%%%%%%%%%%%%%%%%%%%%%%%%%%%%%%%%%%%%%%%
\begin{figure}
\subfloat{\centering\include{Figure6a.tex}}\\
 \subfloat{\centering \includegraphics[width=0.8\columnwidth]{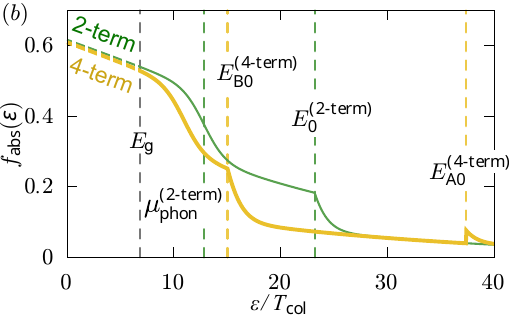}}
  \caption{(a) Sketch of the multi-probe model of the non-equilibrium absorber distribution~\cite{tesser2023}. (b) Electron distribution functions inside the absorber both in the case
  of the two-terminal and four-terminal case. The parameters are $\gamma_\mathrm{colA}=\gamma_\mathrm{colB} = 10\gs$, $E_{\rm B1}=E_{\rm A0}$, $E_{\rm A1}=\infty$, $\Eg=6.8\Tc$, $\gamma_\mathrm{elec} = 0.1\gs$, $\gamma_\mathrm{phon} = 0.27\gs$, $\gamma_\mathrm{rec}=0.01\gs$.
  % I do not see $\Eg$  in the figure.
  }
  \label{fig:neq-distri}
\end{figure}
%%%%%%%%%%%%%%%%%%%%%%%%%%%%%%%%%%%%%%%%

%%%%%%%%%%%%%%%%%%%%%%%%%%%%%%%%%%%%%%%%
\vskip 2mm
\noindent
\textit{Full model: non-equilibrium distribution inside the absorber.---}
To find the distribution function in the absorber, we follow Ref.~\cite{Tesser2022,tesser2023} to model the competing effects sketched in Fig.~\ref{fig:neq-distri}(a). 
Here, electron-hole pair creation  by incident photons is modeled via a Fermi distribution at $\ms=0$ and $\Ts = 6000$\,K. Energy relaxation to the phonons is modeled by a distribution at $T_\mathrm{phon}$, typically given by room temperature (300\,K), and a floating electrochemical potential $\mu_\mathrm{phon}$ guaranteeing particle conservation.  Thermalization between carriers is modeled via a floating probe with electrochemical potential $\mu_\text{elec}$ and temperature $T_\text{elec}$ which adjust to conserve particle- and energy currents. 
Recombination of quasiparticle pairs is described by a probe at $T_\mathrm{rec}=T_\mathrm{phon}$ and $\mu_\mathrm{rec}=0$. Finally, each collector terminal is described by the distributions $f_{\mathrm{col},i}$ with an $\epsilon$-dependent coupling due to the energy-filters, $\gci(\varepsilon)\equiv\gamma_\mathrm{col}D_i(\varepsilon)$. Each terminal's chemical potential is optimized (corresponding to optimizing each terminal's load).
The interplay of these effects typically leads to a nonthermal absorber distribution $f_\mathrm{abs}$.
Fig.~\ref{fig:neq-distri}(b) shows examples of this distribution, showing steps at the positions of the energy filters, with nearly-thermal distributions in between the steps.  The distribution need not decay monotonically, for example, the four-terminal distribution jumps up slightly at $\epsilon=E^\text{(4-term)}_{A0}$.

%%%%%%%%%%%%%%%%%%%%%%%%%%%%%%%%%%%%%%%%
\vskip 2mm
\noindent
\textit{Smooth filters.---}
The more experimentally realistic case of smoothed filters can be modeled by replacing the sharp-step transmission function by a smoothed version $D(\epsilon) = (1+\exp[(\varepsilon - E_0)/w])^{-1}$, where $w$ is the smoothness coefficient.  
To predict the effect of this smoothing, we note that the power output is given by integrals over Fermi functions that are all smooth on the scale of $\Tc$
(the smallest temperature in the model). Thus, we expect that if the energy-filters are smoothed on a scale smaller than $\Tc$, it will have little effect on the integrals, and the power output will be similar to the case with sharp energy-filters. To confirm this expectation, we use a version of our full model with the above broadening of energy filters. 
As expected, in both the two-terminal and four-terminal cases, the results for $w \gtrsim 5 \Tc$ are vastly lower than in the absence of broadening.
However, when $w\lesssim \Tc/2$,
the power outputs are extremely close to those at $w=0$ (typically within a few percent). This means that experimentalists do not need perfectly sharp energy-filters, it is enough to engineer  filters with smoothing $w\lesssim \Tc/2$.

%%%%%%%%%%%%%%%%%%%%%%%%%%%%%%%%%%%%%%%%
\bibliography{solar-cell-filters.bib}
%%%%%%%%%%%%%%%%%%%%%%%%%%%%%%%%%%%%%%%%

\end{document}

%% file: Figure6a.tex
\begin{pgfpicture}
\pgfpathrectangle{\pgfpointorigin}{\pgfqpoint{180.0000bp}{137.0000bp}}
\pgfusepath{use as bounding box}
\begin{pgfscope}
\definecolor{fc}{rgb}{0.0000,0.0000,0.0000}
\pgfsetfillcolor{fc}
\pgftransformshift{\pgfqpoint{12.0023bp}{125.7857bp}}
\pgftransformscale{1.0937}
\pgftext[]{$(a)$}
\end{pgfscope}
\begin{pgfscope}
\definecolor{fc}{rgb}{0.0000,0.0000,0.0000}
\pgfsetfillcolor{fc}
\pgftransformshift{\pgfqpoint{90.0000bp}{49.3479bp}}
\pgftransformscale{1.0937}
\pgftext[]{$\tilde{f}_\text{abs}$}
\end{pgfscope}
\begin{pgfscope}
\definecolor{fc}{rgb}{0.0000,0.0000,0.0000}
\pgfsetfillcolor{fc}
\pgfsetfillopacity{0.0000}
\pgfsetlinewidth{0.5000bp}
\definecolor{sc}{rgb}{0.0000,0.0000,0.0000}
\pgfsetstrokecolor{sc}
\pgfsetmiterjoin
\pgfsetbuttcap
\pgfpathqmoveto{105.5995bp}{47.7879bp}
\pgfpathqcurveto{105.5995bp}{56.4033bp}{98.6154bp}{63.3875bp}{90.0000bp}{63.3875bp}
\pgfpathqcurveto{81.3846bp}{63.3875bp}{74.4005bp}{56.4033bp}{74.4005bp}{47.7879bp}
\pgfpathqcurveto{74.4005bp}{39.1725bp}{81.3846bp}{32.1884bp}{90.0000bp}{32.1884bp}
\pgfpathqcurveto{98.6154bp}{32.1884bp}{105.5995bp}{39.1725bp}{105.5995bp}{47.7879bp}
\pgfpathclose
\pgfusepathqfillstroke
\end{pgfscope}
\begin{pgfscope}
\pgfsetlinewidth{0.5000bp}
\definecolor{sc}{rgb}{0.0000,0.0000,0.0000}
\pgfsetstrokecolor{sc}
\pgfsetmiterjoin
\pgfsetbuttcap
\pgfpathqmoveto{108.9147bp}{39.9532bp}
\pgfpathqlineto{140.2358bp}{26.9796bp}
\pgfusepathqstroke
\end{pgfscope}
\begin{pgfscope}
\definecolor{fc}{rgb}{0.0000,0.0000,0.0000}
\pgfsetfillcolor{fc}
\pgfpathqmoveto{104.5498bp}{41.7612bp}
\pgfpathqlineto{110.7105bp}{41.7612bp}
\pgfpathqcurveto{109.0093bp}{41.0565bp}{108.2014bp}{39.1062bp}{108.9061bp}{37.4049bp}
\pgfpathqlineto{104.5498bp}{41.7612bp}
\pgfpathqlineto{104.5498bp}{41.7612bp}
\pgfpathclose
\pgfusepathqfill
\end{pgfscope}
\begin{pgfscope}
\definecolor{fc}{rgb}{0.0000,0.0000,0.0000}
\pgfsetfillcolor{fc}
\pgfpathqmoveto{108.8191bp}{39.7222bp}
\pgfpathqcurveto{108.8712bp}{39.8808bp}{108.9352bp}{40.0352bp}{109.0104bp}{40.1842bp}
\pgfpathqlineto{108.8191bp}{39.7222bp}
\pgfpathclose
\pgfusepathqfill
\end{pgfscope}
\begin{pgfscope}
\definecolor{fc}{rgb}{0.0000,0.0000,0.0000}
\pgfsetfillcolor{fc}
\pgfpathqmoveto{144.6008bp}{25.1716bp}
\pgfpathqlineto{140.2445bp}{29.5278bp}
\pgfpathqcurveto{140.9492bp}{27.8266bp}{140.1413bp}{25.8762bp}{138.4401bp}{25.1716bp}
\pgfpathqlineto{144.6008bp}{25.1716bp}
\pgfpathqlineto{144.6008bp}{25.1716bp}
\pgfpathclose
\pgfusepathqfill
\end{pgfscope}
\begin{pgfscope}
\definecolor{fc}{rgb}{0.0000,0.0000,0.0000}
\pgfsetfillcolor{fc}
\pgfpathqmoveto{140.1401bp}{26.7486bp}
\pgfpathqcurveto{140.2154bp}{26.8976bp}{140.2794bp}{27.0520bp}{140.3315bp}{27.2105bp}
\pgfpathqlineto{140.1401bp}{26.7486bp}
\pgfpathclose
\pgfusepathqfill
\end{pgfscope}
\begin{pgfscope}
\definecolor{fc}{rgb}{0.0000,0.0000,0.0000}
\pgfsetfillcolor{fc}
\pgftransformshift{\pgfqpoint{122.3471bp}{26.0031bp}}
\pgftransformscale{1.0937}
\pgftext[]{$\gamma_\text{col}$}
\end{pgfscope}
\begin{pgfscope}
\definecolor{fc}{rgb}{0.0000,0.0000,0.0000}
\pgfsetfillcolor{fc}
\pgftransformshift{\pgfqpoint{162.0605bp}{17.9395bp}}
\pgftransformscale{1.0937}
\pgftext[]{\begin{tabular}{c}$T_\text{col}$\\ $\mu_\text{A}$\end{tabular}}
\end{pgfscope}
\begin{pgfscope}
\definecolor{fc}{rgb}{0.0000,0.0000,1.0000}
\pgfsetfillcolor{fc}
\pgfsetfillopacity{0.3000}
\pgfpathqmoveto{180.0000bp}{17.9395bp}
\pgfpathqcurveto{180.0000bp}{27.8472bp}{171.9682bp}{35.8790bp}{162.0605bp}{35.8790bp}
\pgfpathqcurveto{152.1528bp}{35.8790bp}{144.1210bp}{27.8472bp}{144.1210bp}{17.9395bp}
\pgfpathqcurveto{144.1210bp}{8.0318bp}{152.1528bp}{-0.0000bp}{162.0605bp}{-0.0000bp}
\pgfpathqcurveto{171.9682bp}{-0.0000bp}{180.0000bp}{8.0318bp}{180.0000bp}{17.9395bp}
\pgfpathclose
\pgfusepathqfill
\end{pgfscope}
\begin{pgfscope}
\pgfsetlinewidth{0.5000bp}
\definecolor{sc}{rgb}{0.0000,0.0000,0.0000}
\pgfsetstrokecolor{sc}
\pgfsetmiterjoin
\pgfsetbuttcap
\pgfpathqmoveto{71.0853bp}{39.9532bp}
\pgfpathqlineto{39.7642bp}{26.9796bp}
\pgfusepathqstroke
\end{pgfscope}
\begin{pgfscope}
\definecolor{fc}{rgb}{0.0000,0.0000,0.0000}
\pgfsetfillcolor{fc}
\pgfpathqmoveto{75.4502bp}{41.7612bp}
\pgfpathqlineto{71.0939bp}{37.4049bp}
\pgfpathqcurveto{71.7986bp}{39.1062bp}{70.9907bp}{41.0565bp}{69.2895bp}{41.7612bp}
\pgfpathqlineto{75.4502bp}{41.7612bp}
\pgfpathqlineto{75.4502bp}{41.7612bp}
\pgfpathclose
\pgfusepathqfill
\end{pgfscope}
\begin{pgfscope}
\definecolor{fc}{rgb}{0.0000,0.0000,0.0000}
\pgfsetfillcolor{fc}
\pgfpathqmoveto{70.9896bp}{40.1842bp}
\pgfpathqcurveto{71.0648bp}{40.0352bp}{71.1288bp}{39.8808bp}{71.1809bp}{39.7222bp}
\pgfpathqlineto{70.9896bp}{40.1842bp}
\pgfpathclose
\pgfusepathqfill
\end{pgfscope}
\begin{pgfscope}
\definecolor{fc}{rgb}{0.0000,0.0000,0.0000}
\pgfsetfillcolor{fc}
\pgfpathqmoveto{35.3992bp}{25.1716bp}
\pgfpathqlineto{41.5599bp}{25.1716bp}
\pgfpathqcurveto{39.8587bp}{25.8762bp}{39.0508bp}{27.8266bp}{39.7555bp}{29.5278bp}
\pgfpathqlineto{35.3992bp}{25.1716bp}
\pgfpathqlineto{35.3992bp}{25.1716bp}
\pgfpathclose
\pgfusepathqfill
\end{pgfscope}
\begin{pgfscope}
\definecolor{fc}{rgb}{0.0000,0.0000,0.0000}
\pgfsetfillcolor{fc}
\pgfpathqmoveto{39.6685bp}{27.2105bp}
\pgfpathqcurveto{39.7206bp}{27.0520bp}{39.7846bp}{26.8976bp}{39.8599bp}{26.7486bp}
\pgfpathqlineto{39.6685bp}{27.2105bp}
\pgfpathclose
\pgfusepathqfill
\end{pgfscope}
\begin{pgfscope}
\definecolor{fc}{rgb}{0.0000,0.0000,0.0000}
\pgfsetfillcolor{fc}
\pgftransformshift{\pgfqpoint{51.7230bp}{40.3193bp}}
\pgftransformscale{1.0937}
\pgftext[]{$\gamma_\text{col}$}
\end{pgfscope}
\begin{pgfscope}
\definecolor{fc}{rgb}{0.0000,0.0000,0.0000}
\pgfsetfillcolor{fc}
\pgftransformshift{\pgfqpoint{17.9395bp}{17.9395bp}}
\pgftransformscale{1.0937}
\pgftext[]{\begin{tabular}{c}$T_\text{col}$\\ $\mu_\text{B}$\end{tabular}}
\end{pgfscope}
\begin{pgfscope}
\definecolor{fc}{rgb}{0.0000,0.0000,1.0000}
\pgfsetfillcolor{fc}
\pgfsetfillopacity{0.3000}
\pgfpathqmoveto{35.8790bp}{17.9395bp}
\pgfpathqcurveto{35.8790bp}{27.8472bp}{27.8472bp}{35.8790bp}{17.9395bp}{35.8790bp}
\pgfpathqcurveto{8.0318bp}{35.8790bp}{0.0000bp}{27.8472bp}{0.0000bp}{17.9395bp}
\pgfpathqcurveto{0.0000bp}{8.0318bp}{8.0318bp}{0.0000bp}{17.9395bp}{0.0000bp}
\pgfpathqcurveto{27.8472bp}{0.0000bp}{35.8790bp}{8.0318bp}{35.8790bp}{17.9395bp}
\pgfpathclose
\pgfusepathqfill
\end{pgfscope}
\begin{pgfscope}
\pgfsetlinewidth{0.5000bp}
\definecolor{sc}{rgb}{0.0000,0.0000,0.0000}
\pgfsetstrokecolor{sc}
\pgfsetmiterjoin
\pgfsetbuttcap
\pgfpathqmoveto{71.0853bp}{55.6227bp}
\pgfpathqlineto{39.7642bp}{68.5963bp}
\pgfusepathqstroke
\end{pgfscope}
\begin{pgfscope}
\definecolor{fc}{rgb}{0.0000,0.0000,0.0000}
\pgfsetfillcolor{fc}
\pgfpathqmoveto{75.4502bp}{53.8147bp}
\pgfpathqlineto{69.2895bp}{53.8147bp}
\pgfpathqcurveto{70.9907bp}{54.5193bp}{71.7986bp}{56.4697bp}{71.0939bp}{58.1709bp}
\pgfpathqlineto{75.4502bp}{53.8147bp}
\pgfpathqlineto{75.4502bp}{53.8147bp}
\pgfpathclose
\pgfusepathqfill
\end{pgfscope}
\begin{pgfscope}
\definecolor{fc}{rgb}{0.0000,0.0000,0.0000}
\pgfsetfillcolor{fc}
\pgfpathqmoveto{71.1809bp}{55.8536bp}
\pgfpathqcurveto{71.1288bp}{55.6951bp}{71.0648bp}{55.5407bp}{70.9896bp}{55.3917bp}
\pgfpathqlineto{71.1809bp}{55.8536bp}
\pgfpathclose
\pgfusepathqfill
\end{pgfscope}
\begin{pgfscope}
\definecolor{fc}{rgb}{0.0000,0.0000,0.0000}
\pgfsetfillcolor{fc}
\pgfpathqmoveto{35.3992bp}{70.4043bp}
\pgfpathqlineto{39.7555bp}{66.0480bp}
\pgfpathqcurveto{39.0508bp}{67.7493bp}{39.8587bp}{69.6996bp}{41.5599bp}{70.4043bp}
\pgfpathqlineto{35.3992bp}{70.4043bp}
\pgfpathqlineto{35.3992bp}{70.4043bp}
\pgfpathclose
\pgfusepathqfill
\end{pgfscope}
\begin{pgfscope}
\definecolor{fc}{rgb}{0.0000,0.0000,0.0000}
\pgfsetfillcolor{fc}
\pgfpathqmoveto{39.8599bp}{68.8272bp}
\pgfpathqcurveto{39.7846bp}{68.6783bp}{39.7206bp}{68.5239bp}{39.6685bp}{68.3653bp}
\pgfpathqlineto{39.8599bp}{68.8272bp}
\pgfpathclose
\pgfusepathqfill
\end{pgfscope}
\begin{pgfscope}
\definecolor{fc}{rgb}{0.0000,0.0000,0.0000}
\pgfsetfillcolor{fc}
\pgftransformshift{\pgfqpoint{57.6529bp}{69.5728bp}}
\pgftransformscale{1.0937}
\pgftext[]{$\gamma_\text{sun}$}
\end{pgfscope}
\begin{pgfscope}
\definecolor{fc}{rgb}{0.0000,0.0000,0.0000}
\pgfsetfillcolor{fc}
\pgftransformshift{\pgfqpoint{17.9395bp}{77.6364bp}}
\pgftransformscale{1.0937}
\pgftext[]{\begin{tabular}{c}$T_\text{sun}$\\ $0$\end{tabular}}
\end{pgfscope}
\begin{pgfscope}
\definecolor{fc}{rgb}{1.0000,0.0000,0.0000}
\pgfsetfillcolor{fc}
\pgfsetfillopacity{0.3000}
\pgfpathqmoveto{35.8790bp}{77.6364bp}
\pgfpathqcurveto{35.8790bp}{87.5441bp}{27.8472bp}{95.5759bp}{17.9395bp}{95.5759bp}
\pgfpathqcurveto{8.0318bp}{95.5759bp}{0.0000bp}{87.5441bp}{0.0000bp}{77.6364bp}
\pgfpathqcurveto{0.0000bp}{67.7287bp}{8.0318bp}{59.6969bp}{17.9395bp}{59.6969bp}
\pgfpathqcurveto{27.8472bp}{59.6969bp}{35.8790bp}{67.7287bp}{35.8790bp}{77.6364bp}
\pgfpathclose
\pgfusepathqfill
\end{pgfscope}
\begin{pgfscope}
\pgfsetlinewidth{0.5000bp}
\definecolor{sc}{rgb}{0.0000,0.0000,0.0000}
\pgfsetstrokecolor{sc}
\pgfsetmiterjoin
\pgfsetbuttcap
\pgfpathqmoveto{82.1653bp}{66.7027bp}
\pgfpathqlineto{69.1916bp}{98.0237bp}
\pgfusepathqstroke
\end{pgfscope}
\begin{pgfscope}
\definecolor{fc}{rgb}{0.0000,0.0000,0.0000}
\pgfsetfillcolor{fc}
\pgfpathqmoveto{83.9733bp}{62.3377bp}
\pgfpathqlineto{79.6170bp}{66.6940bp}
\pgfpathqcurveto{81.3182bp}{65.9893bp}{83.2686bp}{66.7972bp}{83.9733bp}{68.4984bp}
\pgfpathqlineto{83.9733bp}{62.3377bp}
\pgfpathqlineto{83.9733bp}{62.3377bp}
\pgfpathclose
\pgfusepathqfill
\end{pgfscope}
\begin{pgfscope}
\definecolor{fc}{rgb}{0.0000,0.0000,0.0000}
\pgfsetfillcolor{fc}
\pgfpathqmoveto{82.3962bp}{66.7983bp}
\pgfpathqcurveto{82.2473bp}{66.7231bp}{82.0928bp}{66.6591bp}{81.9343bp}{66.6070bp}
\pgfpathqlineto{82.3962bp}{66.7983bp}
\pgfpathclose
\pgfusepathqfill
\end{pgfscope}
\begin{pgfscope}
\definecolor{fc}{rgb}{0.0000,0.0000,0.0000}
\pgfsetfillcolor{fc}
\pgfpathqmoveto{67.3836bp}{102.3887bp}
\pgfpathqlineto{67.3836bp}{96.2280bp}
\pgfpathqcurveto{68.0883bp}{97.9292bp}{70.0387bp}{98.7371bp}{71.7399bp}{98.0324bp}
\pgfpathqlineto{67.3836bp}{102.3887bp}
\pgfpathqlineto{67.3836bp}{102.3887bp}
\pgfpathclose
\pgfusepathqfill
\end{pgfscope}
\begin{pgfscope}
\definecolor{fc}{rgb}{0.0000,0.0000,0.0000}
\pgfsetfillcolor{fc}
\pgfpathqmoveto{69.4226bp}{98.1194bp}
\pgfpathqcurveto{69.2641bp}{98.0673bp}{69.1096bp}{98.0033bp}{68.9607bp}{97.9281bp}
\pgfpathqlineto{69.4226bp}{98.1194bp}
\pgfpathclose
\pgfusepathqfill
\end{pgfscope}
\begin{pgfscope}
\definecolor{fc}{rgb}{0.0000,0.0000,0.0000}
\pgfsetfillcolor{fc}
\pgftransformshift{\pgfqpoint{84.8713bp}{86.0650bp}}
\pgftransformscale{1.0937}
\pgftext[]{$\gamma_\text{elec}$}
\end{pgfscope}
\begin{pgfscope}
\definecolor{fc}{rgb}{0.0000,0.0000,0.0000}
\pgfsetfillcolor{fc}
\pgftransformshift{\pgfqpoint{60.1516bp}{119.8484bp}}
\pgftransformscale{1.0937}
\pgftext[]{\begin{tabular}{c}$T_\text{elec}$\\ $\mu_\text{e}$\end{tabular}}
\end{pgfscope}
\begin{pgfscope}
\definecolor{fc}{rgb}{0.0000,0.5020,0.0000}
\pgfsetfillcolor{fc}
\pgfsetfillopacity{0.3000}
\pgfpathqmoveto{78.0910bp}{119.8484bp}
\pgfpathqcurveto{78.0910bp}{129.7561bp}{70.0593bp}{137.7879bp}{60.1516bp}{137.7879bp}
\pgfpathqcurveto{50.2439bp}{137.7879bp}{42.2121bp}{129.7561bp}{42.2121bp}{119.8484bp}
\pgfpathqcurveto{42.2121bp}{109.9407bp}{50.2439bp}{101.9090bp}{60.1516bp}{101.9090bp}
\pgfpathqcurveto{70.0593bp}{101.9090bp}{78.0910bp}{109.9407bp}{78.0910bp}{119.8484bp}
\pgfpathclose
\pgfusepathqfill
\end{pgfscope}
\begin{pgfscope}
\pgfsetlinewidth{0.5000bp}
\definecolor{sc}{rgb}{0.0000,0.0000,0.0000}
\pgfsetstrokecolor{sc}
\pgfsetmiterjoin
\pgfsetbuttcap
\pgfpathqmoveto{97.8347bp}{66.7027bp}
\pgfpathqlineto{110.8084bp}{98.0237bp}
\pgfusepathqstroke
\end{pgfscope}
\begin{pgfscope}
\definecolor{fc}{rgb}{0.0000,0.0000,0.0000}
\pgfsetfillcolor{fc}
\pgfpathqmoveto{96.0267bp}{62.3377bp}
\pgfpathqlineto{96.0267bp}{68.4984bp}
\pgfpathqcurveto{96.7314bp}{66.7972bp}{98.6818bp}{65.9893bp}{100.3830bp}{66.6940bp}
\pgfpathqlineto{96.0267bp}{62.3377bp}
\pgfpathqlineto{96.0267bp}{62.3377bp}
\pgfpathclose
\pgfusepathqfill
\end{pgfscope}
\begin{pgfscope}
\definecolor{fc}{rgb}{0.0000,0.0000,0.0000}
\pgfsetfillcolor{fc}
\pgfpathqmoveto{98.0657bp}{66.6070bp}
\pgfpathqcurveto{97.9072bp}{66.6591bp}{97.7527bp}{66.7231bp}{97.6038bp}{66.7983bp}
\pgfpathqlineto{98.0657bp}{66.6070bp}
\pgfpathclose
\pgfusepathqfill
\end{pgfscope}
\begin{pgfscope}
\definecolor{fc}{rgb}{0.0000,0.0000,0.0000}
\pgfsetfillcolor{fc}
\pgfpathqmoveto{112.6164bp}{102.3887bp}
\pgfpathqlineto{108.2601bp}{98.0324bp}
\pgfpathqcurveto{109.9613bp}{98.7371bp}{111.9117bp}{97.9292bp}{112.6164bp}{96.2280bp}
\pgfpathqlineto{112.6164bp}{102.3887bp}
\pgfpathqlineto{112.6164bp}{102.3887bp}
\pgfpathclose
\pgfusepathqfill
\end{pgfscope}
\begin{pgfscope}
\definecolor{fc}{rgb}{0.0000,0.0000,0.0000}
\pgfsetfillcolor{fc}
\pgfpathqmoveto{111.0393bp}{97.9281bp}
\pgfpathqcurveto{110.8904bp}{98.0033bp}{110.7359bp}{98.0673bp}{110.5774bp}{98.1194bp}
\pgfpathqlineto{111.0393bp}{97.9281bp}
\pgfpathclose
\pgfusepathqfill
\end{pgfscope}
\begin{pgfscope}
\definecolor{fc}{rgb}{0.0000,0.0000,0.0000}
\pgfsetfillcolor{fc}
\pgftransformshift{\pgfqpoint{117.2447bp}{80.1350bp}}
\pgftransformscale{1.0937}
\pgftext[]{$\gamma_\text{phon}$}
\end{pgfscope}
\begin{pgfscope}
\definecolor{fc}{rgb}{0.0000,0.0000,0.0000}
\pgfsetfillcolor{fc}
\pgftransformshift{\pgfqpoint{119.8484bp}{119.8484bp}}
\pgftransformscale{1.0937}
\pgftext[]{\begin{tabular}{c}$T_\text{phon}$\\ $\mu_\text{p}$\end{tabular}}
\end{pgfscope}
\begin{pgfscope}
\definecolor{fc}{rgb}{0.0000,0.0000,1.0000}
\pgfsetfillcolor{fc}
\pgfsetfillopacity{0.3000}
\pgfpathqmoveto{137.7879bp}{119.8484bp}
\pgfpathqcurveto{137.7879bp}{129.7561bp}{129.7561bp}{137.7879bp}{119.8484bp}{137.7879bp}
\pgfpathqcurveto{109.9407bp}{137.7879bp}{101.9090bp}{129.7561bp}{101.9090bp}{119.8484bp}
\pgfpathqcurveto{101.9090bp}{109.9407bp}{109.9407bp}{101.9090bp}{119.8484bp}{101.9090bp}
\pgfpathqcurveto{129.7561bp}{101.9090bp}{137.7879bp}{109.9407bp}{137.7879bp}{119.8484bp}
\pgfpathclose
\pgfusepathqfill
\end{pgfscope}
\begin{pgfscope}
\pgfsetlinewidth{0.5000bp}
\definecolor{sc}{rgb}{0.0000,0.0000,0.0000}
\pgfsetstrokecolor{sc}
\pgfsetmiterjoin
\pgfsetbuttcap
\pgfpathqmoveto{108.9147bp}{55.6227bp}
\pgfpathqlineto{140.2358bp}{68.5963bp}
\pgfusepathqstroke
\end{pgfscope}
\begin{pgfscope}
\definecolor{fc}{rgb}{0.0000,0.0000,0.0000}
\pgfsetfillcolor{fc}
\pgfpathqmoveto{104.5498bp}{53.8147bp}
\pgfpathqlineto{108.9061bp}{58.1709bp}
\pgfpathqcurveto{108.2014bp}{56.4697bp}{109.0093bp}{54.5193bp}{110.7105bp}{53.8147bp}
\pgfpathqlineto{104.5498bp}{53.8147bp}
\pgfpathqlineto{104.5498bp}{53.8147bp}
\pgfpathclose
\pgfusepathqfill
\end{pgfscope}
\begin{pgfscope}
\definecolor{fc}{rgb}{0.0000,0.0000,0.0000}
\pgfsetfillcolor{fc}
\pgfpathqmoveto{109.0104bp}{55.3917bp}
\pgfpathqcurveto{108.9352bp}{55.5407bp}{108.8712bp}{55.6951bp}{108.8191bp}{55.8536bp}
\pgfpathqlineto{109.0104bp}{55.3917bp}
\pgfpathclose
\pgfusepathqfill
\end{pgfscope}
\begin{pgfscope}
\definecolor{fc}{rgb}{0.0000,0.0000,0.0000}
\pgfsetfillcolor{fc}
\pgfpathqmoveto{144.6008bp}{70.4043bp}
\pgfpathqlineto{138.4401bp}{70.4043bp}
\pgfpathqcurveto{140.1413bp}{69.6996bp}{140.9492bp}{67.7493bp}{140.2445bp}{66.0480bp}
\pgfpathqlineto{144.6008bp}{70.4043bp}
\pgfpathqlineto{144.6008bp}{70.4043bp}
\pgfpathclose
\pgfusepathqfill
\end{pgfscope}
\begin{pgfscope}
\definecolor{fc}{rgb}{0.0000,0.0000,0.0000}
\pgfsetfillcolor{fc}
\pgfpathqmoveto{140.3315bp}{68.3653bp}
\pgfpathqcurveto{140.2794bp}{68.5239bp}{140.2154bp}{68.6783bp}{140.1401bp}{68.8272bp}
\pgfpathqlineto{140.3315bp}{68.3653bp}
\pgfpathclose
\pgfusepathqfill
\end{pgfscope}
\begin{pgfscope}
\definecolor{fc}{rgb}{0.0000,0.0000,0.0000}
\pgfsetfillcolor{fc}
\pgftransformshift{\pgfqpoint{128.2770bp}{55.2565bp}}
\pgftransformscale{1.0937}
\pgftext[]{$\gamma_\text{rec}$}
\end{pgfscope}
\begin{pgfscope}
\definecolor{fc}{rgb}{0.0000,0.0000,0.0000}
\pgfsetfillcolor{fc}
\pgftransformshift{\pgfqpoint{162.0605bp}{77.6364bp}}
\pgftransformscale{1.0937}
\pgftext[]{\begin{tabular}{c}$T_\text{rec}$\\ $0$\end{tabular}}
\end{pgfscope}
\begin{pgfscope}
\definecolor{fc}{rgb}{0.0000,0.0000,1.0000}
\pgfsetfillcolor{fc}
\pgfsetfillopacity{0.3000}
\pgfpathqmoveto{180.0000bp}{77.6364bp}
\pgfpathqcurveto{180.0000bp}{87.5441bp}{171.9682bp}{95.5759bp}{162.0605bp}{95.5759bp}
\pgfpathqcurveto{152.1528bp}{95.5759bp}{144.1210bp}{87.5441bp}{144.1210bp}{77.6364bp}
\pgfpathqcurveto{144.1210bp}{67.7287bp}{152.1528bp}{59.6969bp}{162.0605bp}{59.6969bp}
\pgfpathqcurveto{171.9682bp}{59.6969bp}{180.0000bp}{67.7287bp}{180.0000bp}{77.6364bp}
\pgfpathclose
\pgfusepathqfill
\end{pgfscope}
\end{pgfpicture}